\documentclass[12pt]{article}
\usepackage{graphicx}

\def \b{{\cal B}}
\def \bea{\begin{eqnarray}}
\def \beq{\begin{equation}}

\def \eea{\end{eqnarray}}
\def \eeq{\end{equation}}

\def \ok{\overline{K}^0}
\def \ol{\overline}
\def \s{\sqrt{2}}
\def \st{\sqrt{3}}
\def \sx{\sqrt{6}}

\begin{document}
\rightline{EFI 08-32}
\rightline{arXiv:yymm.nnnn}
\rightline{December 2008}
\bigskip
\centerline{\bf DECAYS OF CHARMED MESONS TO $PV$ FINAL STATES}
\bigskip

\centerline{Bhubanjyoti Bhattacharya\footnote{bhujyo@uchicago.edu} and
Jonathan L. Rosner\footnote{rosner@hep.uchicago.edu}}
\centerline{\it Enrico Fermi Institute and Department of Physics}
\centerline{\it University of Chicago, 5640 S. Ellis Avenue, Chicago, IL 60637}

\begin{quote}

New data on the decays of the charmed particles $D^0$, $D^+$, and $D_s$ to
$PV$ final states consisting of a light pseudoscalar meson $P$ and a light
vector meson $V$ are analyzed.  Following the same methods as in a previous
analysis of $D \to PP$ decays, one can test flavor symmetry, extract key
key amplitudes, and obtain information on relative strong phases.  Analyses are
performed for Cabibbo-favored decays and then extended to predict properties of
singly- and doubly-Cabibbo-suppressed processes.
\end{quote}

\section{INTRODUCTION}

In the past few years rich data on charmed particle decays have been
contributed by a variety of experiments.  Among the decays studied are those
involving $PV$ final states, where $P$ and $V$ denote light pseudoscalar and
vector mesons, respectively.  These decays obey an approximate flavor
SU(3) symmetry \cite{Chau:1983,Chau:1986,Chiang:2002mr}, allowing one to
investigate such questions as the strong phases of amplitudes in these decays.
These strong phases can be important when analyzing $D$ decay Dalitz plots
in the context of studies of CP violation in $B \to D X$ decays.  We have
recently performed a similar analysis of $D \to PP$ decays
\cite{Bhattacharya:2008ss}.

The diagrammatic approach to flavor symmetry is reviewed briefly in Section II.
Cabibbo-favored decays are discussed in Section III, singly-Cabibbo-suppressed
decays in Section IV, and doubly-Cabibbo-suppressed decays in Section V. It is
possible to obtain a few of the relevant amplitudes using factorization
techniques.  We discuss factorization calculations in Section VI and conclude
in Section VII.

\section{DIAGRAMMATIC AMPLITUDE EXPANSION}

A flavor-topology description of $D \to PV$ decays uses amplitudes defined
as in Ref.\ \cite{Chiang:2002mr}.  Cabibbo-favored (CF) amplitudes,
proportional to the product $V_{ud} V^*_{cs}$ of Cabibbo-Kobayashi-Maskawa
(CKM) factors, will be denoted by unprimed quantities;
singly-Cabibbo-suppressed amplitudes proportional to $V_{us} V^*_{cs}$ or
$V_{ud} V^*_{cd}$ will be denoted by primed quantities; and
doubly-Cabibbo-suppressed quantities proportional to $V_{us} V^*_{cd}$ will
be denoted by amplitudes with a tilde.  These amplitudes are in the ratio
$1:\lambda:-\lambda:-\lambda^2$, where $\lambda = \tan \theta_C =
0.2317$ \cite{Amsler:2008}, with $\theta_C$ the Cabibbo angle.

The relevant amplitudes are labeled as $T$ (``tree''), $C$
(``color-suppressed''), $E$ (``exchange''), and (``A'') (annihilation).
For $PV$ final states, a subscript on the amplitude denotes the meson
($P$ or $V$) containing the spectator quark.

The partial width $\Gamma(H \to PV)$ for the decay of a heavy meson $H$ may
be expressed in terms of an invariant amplitude ${\cal A}$ as
\beq
\Gamma(H \to PV) = \frac{p^{*3}}{8 \pi M_H^2}|{\cal A}|^2~,
\eeq
where $p^*$ is the center-of-mass (c.m.) 3-momentum of each final particle,
and $M_H$ is the mass of the decaying particle.

\section{CABIBBO-FAVORED DECAYS}

\begin{figure}
\begin{center}
\mbox{\includegraphics[width=0.51\textwidth]{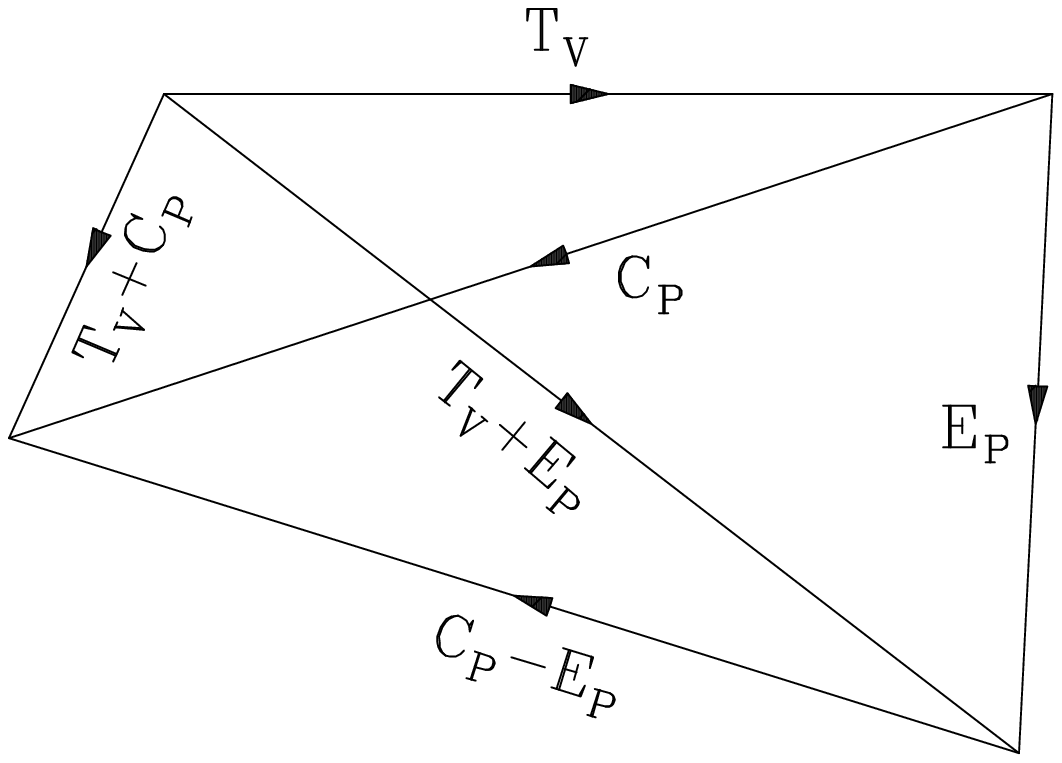} \hskip 0.3in
      \includegraphics[width=0.43\textwidth]{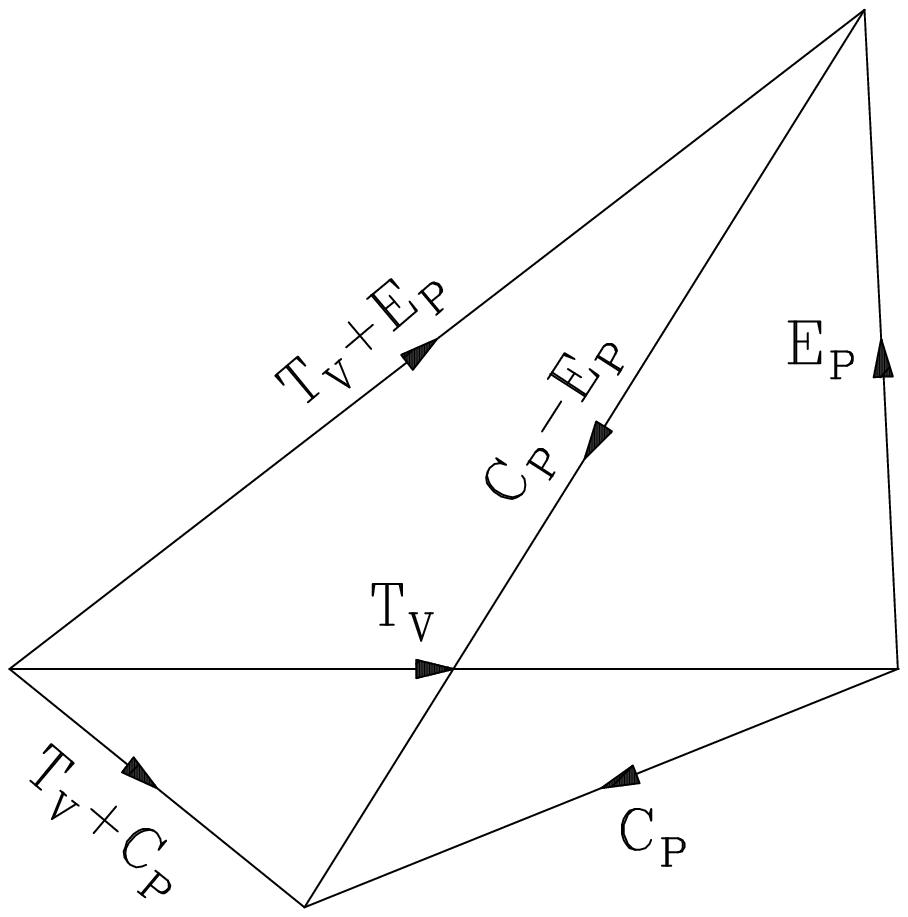}}
\end{center}
\caption{Magnitudes of and relative phases between $T_P$, $C_V$ and $E_V$.
Left: solution (``A'') with $|C_P| > |T_V|$; right: solution (``B'') with
$|C_P| < |T_V|$.
\label{fig:TCEamps}}
\end{figure}

In Table \ref{tab:CFPV} we summarize predicted and observed amplitudes for
Cabibbo-favored decays of charmed mesons to $PV$.  The experimental values
are based on those in Ref.\ \cite{Amsler:2008} unless noted otherwise.
Topological amplitudes are then obtained from these processes by algebraic
solution.  The values of $|T_V|$ and $|E_P|$ are uniquely
given by the rates for $D_s \to \pi^+ \phi$ and $D^0 \to \ol{K}^{0} \phi$,
respectively. A two-fold ambiguity then is found for the amplitude $|C_P|$ and
phases of $C_P$ and $E_P$, as summarized in Table \ref{tab:cfampsa}.

As explained in Ref.\ \cite{Rosner:1999}, the solution ``B'' with $|C_P| <
|T_V|$ is expected for a color suppressed amplitude.  However, on the basis of
fits to data from singly-Cabibbo-suppressed $D \to PV$ decays, it will turn out
that we will prefer the solution ``A'' with $|C_P| > |T_V|$.  In Fig.\
\ref{fig:TCEamps} we plot these two solutions for amplitudes and relative
phases of $T_V$, $C_P$ and $E_P$.
\newpage

\begin{table}
\caption{Branching ratios and invariant amplitudes for Cabibbo-favored
decays of charmed mesons to one pseudoscalar and one vector meson.
\label{tab:CFPV}}
\begin{center}
\begin{tabular}{c l c c c c}
\hline \hline
Meson & Decay & Representation
     & ${\cal B}$ \cite{Amsler:2008} & $p^*$ & $|{\cal A}|$ \\
 & mode & & ($\%$) & (MeV) & $(10^{-6})$ \\ \hline \hline
$D^0$ & $K^{*-} \pi^+$ & $T_V + E_P$
     & $5.91 \pm 0.39$ & 710.9 & $4.80 \pm 0.16$ \\
 & $K^- \rho^+$ & $T_P + E_V$
     & $10.8 \pm 0.7$ & 675.4 & $7.01 \pm 0.23$ \\
 & $\ol{K}^{*0} \pi^0$ & $\frac{1}{\sqrt{2}}(C_P - E_P)$
     & $2.82 \pm 0.35$ & 709.3 & $3.33 \pm 0.21$ \\
 & $\ol{K}^0 \rho^0$ & $\frac{1}{\sqrt{2}}(C_V - E_V)$
     & $1.54 \pm 0.12$ & 673.7 & $2.66 \pm 0.14$ \\
 & $\ol{K}^{*0} \eta$ & $\frac{1}{\sqrt{3}}(C_P + E_P - E_V)$
     & $0.96 \pm 0.3$ & 579.9 & $2.63 \pm 0.41$ \\
 & $\ol{K}^{*0} \eta\,'$ & $-\frac{1}{\sqrt{6}}(C_P + E_P + 2 E_V)$
     & $< 0.11$& 101.9 & \\
 & $\ol{K}^0 \omega$ & $-\frac{1}{\sqrt{2}}(C_V + E_V)$
     & $2.26 \pm 0.4$ & 670.0 & $3.25 \pm 0.29$ \\
 & $\ol{K}^0 \phi$ & $-E_P$
     & $0.868 \pm 0.06$ & 520.6 & $2.94 \pm 0.10$ \\
\hline
$D^+$ & $\ol{K}^{*0} \pi^+$ & $T_V + C_P$
     & $1.83 \pm 0.14$ & 711.8 & $1.68 \pm 0.06$ \\
 & $\ol{K}^0 \rho^+$ & $T_P + C_V$
     & $9.2 \pm 2.0$ & 677.0 & $4.06 \pm 0.44$ \\
\hline
$D_s^+$ & $\ol{K}^{*0} K^+$ & $C_P + A_V$
     & $3.9 \pm 0.6$ & 682.4 & $3.97 \pm 0.31$ \\
 & $\ol{K}^0 K^{*+}$ & $C_V + A_P$
     & $5.3 \pm 1.2$ & 683.2 & $4.61 \pm 0.52$ \\
 & $\rho^+ \pi^0$ & $\frac{1}{\sqrt{2}}(A_P - A_V)$ & & 825.2 & \\
 & $\rho^+ \eta$ & $\frac{1}{\sqrt{3}}(T_P - A_P - A_V)$
     & $13.0 \pm 2.2$ & 723.8 & $6.63 \pm 0.56$ \\
 & $\rho^+ \eta\,'$ & $\frac{1}{\sqrt{6}}(2T_P + A_P + A_V)$
     & $12.2 \pm 2.0$  & 464.8 & $12.5 \pm 1.0$ \\
 & $\pi^+ \rho^0$ & $\frac{1}{\sqrt{2}}(A_V - A_P)$
     & & 824.7 & \\
 & $\pi^+ \omega$ & $\frac{1}{\sqrt{2}}(A_V + A_P)$
     & $0.25 \pm 0.09$ & 821.8 & $0.76 \pm 0.14$ \\
 & $\pi^+ \phi$ & $T_V$
     & $4.38 \pm 0.35$ & 711.7 & $3.95 \pm 0.16$ \\
\hline \hline
\end{tabular}
\end{center}
\end{table}

\begin{table}
\caption{Solutions in Cabibbo-favored charmed meson
decays to $PV$ final states.
\label{tab:cfampsa}}
\begin{center}
\begin{tabular}{c c c c c} \hline \hline
          & \multicolumn{2}{c}{Solution A} & \multicolumn{2}{c}{Solution B} \\
   $PV$   &  Magnitude  &   Relative   &  Magnitude  &   Relative   \\
amplitude & ($10^{-6}$) & strong phase & ($10^{-6}$) & strong phase \\ \hline
$T_V$ & $3.95 \pm 0.07$ & --- & $3.95 \pm 0.07$ & --- \\
$C_P$ & $4.88 \pm 0.15$ & $\delta_{C_PT_V} = (-162 \pm 1)^\circ$ & $2.84 \pm 0.09$ & $\delta_{C_PT_V} = (-158.2^{+2.0}_{-1.9})^\circ$\\
$E_P$ & $2.94 \pm 0.09$ & $\delta_{E_PT_V} = (-93  \pm 3)^\circ$ & $2.94 \pm 0.10$ & $\delta_{E_PT_V} = (92.8^{+3.6}_{-3.7})^\circ$\\
\hline\hline
\end{tabular}
\end{center}
\end{table}
\newpage

Using the solutions for $T_V$, $C_P$ and $E_P$ as inputs, the other amplitudes
$T_P$, $C_V$ and $E_V$ were obtained.  The amplitude $T_P$ was assumed real
relative to $T_V$, in accord with the expectation from factorization. Six sets
of solutions were obtained for each of the two cases $|T_V| < |C_P|$ (``A'')
and $|T_V| > |C_P|$ (``B''). These solutions are listed in Table
\ref{tab:cfampsc}. The solutions A1 and A2 are found to give the best fit to
the data available for singly-Cabibbo-suppressed $D \to PV$ decays, and so will
be singled out for special consideration.  Note the identical magnitudes and
phases of $T_P$, $C_V$ and $E_V$ in Solutions A1 and B1.

\begin{figure}
\begin{center}
\includegraphics[width=0.84\textwidth]{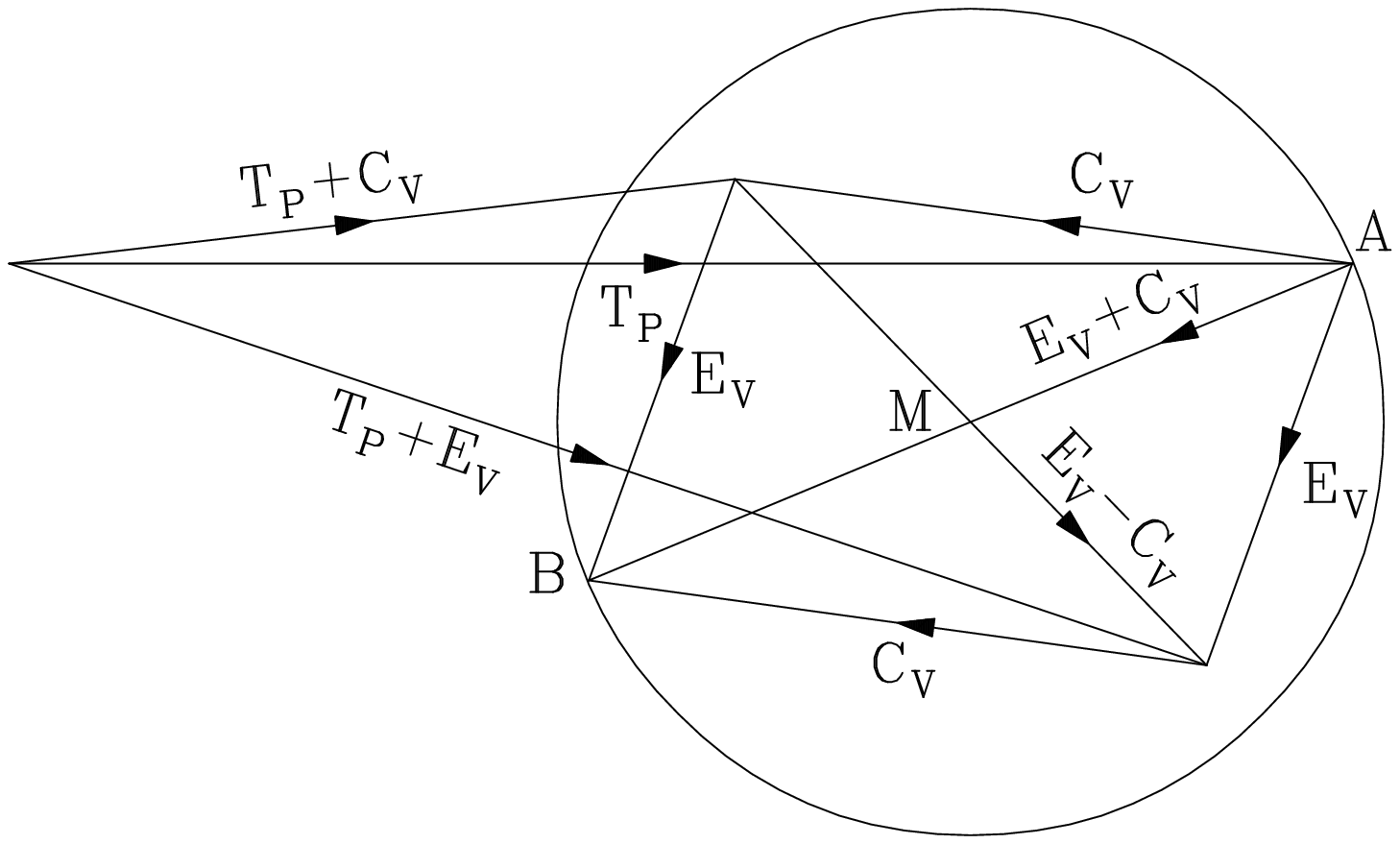}
\includegraphics[width=0.80\textwidth]{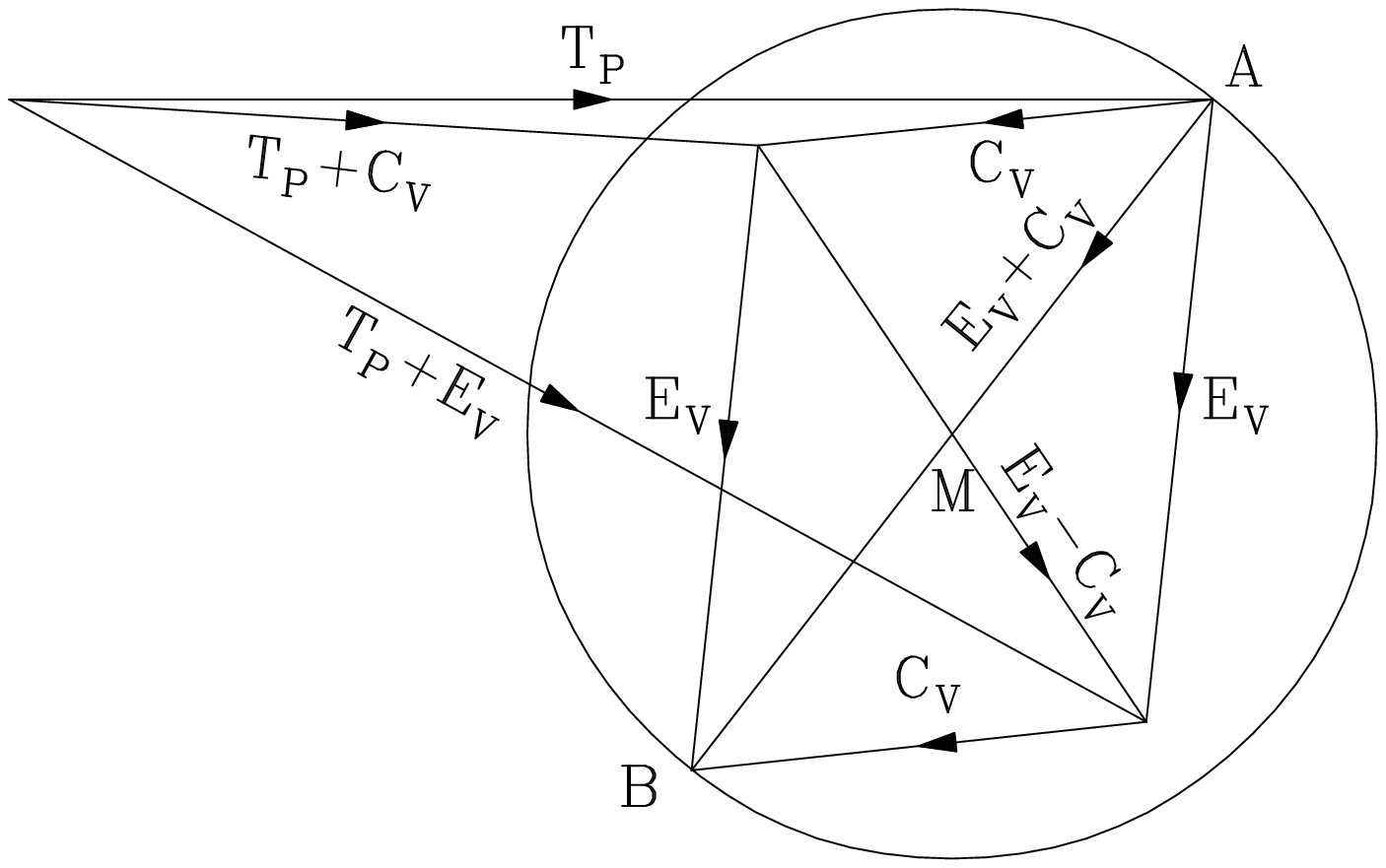}
\end{center}
\caption{Amplitudes $T_P$, $C_V$, and $E_V$ in solutions A1 (top) and A2
(bottom).
\label{fig:cftp}}
\end{figure}

The magnitudes and phases of solutions A1 and A2 are illustrated in
Fig.\ \ref{fig:cftp}.  The amplitudes $T_P + E_V = {\cal A}(D^0 \to K^-
\rho^+)$, $C_V - E_V = \sqrt{2}{\cal A}(D^0 \to \ok \rho^0)$, and $T_P + C_V =
{\cal A}(D^+ \to \ok \rho^+)$ form a triangle whose shape is specified by their
magnitudes.  The amplitudes $C_V$ and $E_V$ form the sides of a quadrangle
whose diagonals are $C_V - E_V = \sqrt{2}{\cal A}(D^0 \to \ok \rho)$ and $C_V +
E_V = -\sqrt{2}{\cal A}(D^0 \to \ok \omega)$, and whose vertices lie on
a circle with midpoint $M$.  Two vertices are fixed, while the other two ($A$
and $B$ in Fig.\ \ref{fig:cftp}) lie at any two opposite points on the circle.
An additional constraint is the magnitude of $C_P + E_P - E_V = \sqrt{3}{\cal
A}(D^0 \to \overline{K}^{*0} \eta)$.  A discrete ambiguity remains,
corresponding to the solutions listed in Tables \ref{tab:cfampsa} and
\ref{tab:cfampsc}.

\begin{table}
\caption{Alternative solutions for $T_P$, $C_V$, and $E_V$ amplitudes in
Cabibbo-favored charmed meson decays to $PV$ final states. Solutions A1 --
A6 correspond to $|T_V| < |C_P|$, while the solutions B1 -- B6 correspond to
$|T_V| > |C_P|$
\label{tab:cfampsc}}
\begin{center}
\begin{tabular}{c c c c c} \hline \hline
   No.  & $PV$  &  Magnitude  &   Relative   & ${\cal{B}}(D^0 \to \ol{K^{*0}}\,\eta\,')$\\
        & ampl. & ($10^{-6}$) &    phase     & ($10^{-4}$) \\ \hline
 A1$^a$ & $T_P$ & 7.46$\pm$0.21 & Assumed 0 & \\
        & $C_V$ & 3.46$\pm$0.18 &$\delta_{C_VT_V} = (172 \pm 3)^\circ$  & $1.52 \pm 0.22$\\
        & $E_V$ & 2.37$\pm$0.19 &$\delta_{E_VT_V} = (-110 \pm 4)^\circ$ & \\ \hline
 A2$^b$ & $T_P$ & 6.51$\pm$0.23 & Assumed 0 & \\
        & $C_V$ & 2.47$\pm$0.22 &$\delta_{C_VT_P} = (-174 \pm 4)^\circ$ & $1.96 \pm 0.23$\\
        & $E_V$ & 3.39$\pm$0.16 &$\delta_{E_VT_P} = (-96 \pm 3)^\circ$  & \\ \hline
   A3   & $T_P$ &--5.67$\pm$0.22& Assumed 0 & \\
        & $C_V$ & 3.64$\pm$0.27 &$\delta_{C_VT_P} = (-46 \pm 4)^\circ$  & $1.42 \pm 0.28$ \\
        & $E_V$ & 2.09$\pm$0.28 &$\delta_{E_VT_P} = (-122^{+5}_{-6})^\circ$ & \\ \hline
   A4   & $T_P$ &--5.60$\pm$0.24& Assumed 0 & \\
        & $C_V$ & 1.68$\pm$0.24 &$\delta_{C_VT_P} = (-20 \pm 6)^\circ$  & $2.21 \pm 0.25$ \\
        & $E_V$ & 3.85$\pm$0.15 &$\delta_{E_VT_P} = (-94 \pm 3)^\circ$  & \\ \hline
   A5   & $T_P$ &--3.22$\pm$0.21& Assumed 0 & \\
        & $C_V$ & 1.79$\pm$0.32 &$\delta_{C_VT_P} = (-104 \pm 5)^\circ$ & $2.18 \pm 0.25$ \\
        & $E_V$ & 3.79$\pm$0.13 &$\delta_{E_VT_P} = (-180^{+4}_{-5})^\circ$ & \\ \hline
   A6   & $T_P$ & 3.21$\pm$0.21 & Assumed 0 & \\
        & $C_V$ & 1.78$\pm$0.31 &$\delta_{C_VT_P} = (105 \pm 5)^\circ$  & $2.18 \pm 0.25$ \\
        & $E_V$ & 3.80$\pm$0.13 &$\delta_{E_VT_P} = (-180^{+5}_{-4})^\circ$ & \\  \hline\hline
   B1   & $T_P$ & 7.46$\pm$0.21 & Assumed 0 & \\
        & $C_V$ & 3.46$\pm$0.17 &$\delta_{C_VT_P} = (172 \pm 3)^\circ$  & 0.33$\pm$0.05 \\
        & $E_V$ & 2.37$\pm$0.19 &$\delta_{E_VT_P} = (-110 \pm 4)^\circ$       & \\ \hline
   B2   & $T_P$ & 6.43$\pm$0.22 & Assumed 0 & \\
        & $C_V$ & 3.95$\pm$0.24 &$\delta_{C_VT_P}=(-143 \pm 4)^\circ$ & $0.052^{+0.020}_{-0.021}$ \\
        & $E_V$ & 1.40$\pm$0.32 &$\delta_{E_VT_P}= (-71^{+6}_{-7})^\circ$   & \\ \hline
   B3   & $T_P$ & 4.53$\pm$0.24 & Assumed 0 & \\
        & $C_V$ & 0.80$\pm$0.21 &$\delta_{C_VT_P}=(130^{+16}_{-15})^\circ$      & 1.18$\pm$0.10 \\
        & $E_V$ & 4.12$\pm$0.15 &$\delta_{E_VT_P}=(72 \pm 3)^\circ$   & \\ \hline
   B4   & $T_P$ & 4.97$\pm$0.22 & Assumed 0 & \\
        & $C_V$ & 3.28$\pm$0.29 &$\delta_{C_VT_P}=(126 \pm 4)^\circ$  & 0.42$\pm$0.10 \\
        & $E_V$ & 2.61$\pm$0.25 &$\delta_{E_VT_P}=(47 \pm 5)^\circ$   & \\ \hline
   B5   & $T_P$ &--3.33$\pm$0.22& Assumed 0 & \\
        & $C_V$ & 0.75$\pm$0.19 &$\delta_{C_VT_V}=(164^{+14}_{-15})^\circ$    & 1.19$\pm$0.11 \\
        & $E_V$ &4.13$\pm$0.17  &$\delta_{E_VT_V}=(-140 \pm 2)^\circ$ & \\ \hline
   B6   & $T_P$ &--7.70$\pm$0.21 & Assumed 0 & \\
        & $C_V$ & 4.01$\pm$0.17 &$\delta_{C_VT_V}=(17^{+3}_{-4})^\circ$   & 0.020$\pm$0.011 \\
        & $E_V$ & 1.24$\pm$0.22 &$\delta_{E_VT_V}=(-52^{+9}_{-8})^\circ$      & \\ \hline\hline
\end{tabular}
\end{center}
\leftline{$^a$Preferred solution based on fit to singly-Cabibbo-suppressed
decays.}
\leftline{$^b$Alternative solution giving acceptable fit to
singly-Cabibbo-suppressed decays.}
\end{table}

Predictions for the
branching ratio for $D^0 \to \overline{K}^{*0} \eta'$, listed in the last
column of Table \ref{tab:cfampsd}, in principle allow one to distinguish among
various solutions.  In addition, we shall see that only solutions A1 and A2
give rise to acceptable fits to singly-Cabibbo-suppressed decays.

We now state a relationship between $|T_P|$ and Cabibbo-favored $D_s$ decay
amplitudes:
\beq
|A(D_s \to \rho^+ \eta')|^2 = |T_P|^2 + |A(D_s \to \pi^+ \omega)|^2
 - |A(D_s \to \rho^+ \eta)|^2
\eeq
Using the value of $|T_P|$ from solution A1 of Table \ref{tab:cfampsc} and the
decay amplitudes $(D_s \to \rho^+ \eta,\ \pi^+ \omega)$ from Table
\ref{tab:CFPV}, we calculate the amplitude: $|A(D_s \to \rho^+ \eta')| =
(3.50 \pm 1.15) \times 10^{-6}$, which deviates from the experimental value
(Table \ref{tab:CFPV}) by a large amount.  This problem with the quoted
experimental rate for $D_s \to \rho^+ \eta'$ was already noted in Ref.\
\cite{Rosner:1999}. It indicates either the importance of neglected amplitudes
involving the flavor-singlet component of $\eta'$, or an overestimate of the
experimental decay rate in this mode.

The remaining parameters $A_P$ and $A_V$ were determined using the amplitudes
of $D_s \to (\ol{K}^{*0} K^+,\ \ol{K}^0 K^{*+},\ \pi^+ \omega)$ and have been
listed in Table \ref{tab:cfampsd}. A direct calculation of the amplitudes for
$D_s \to \rho^+ (\eta,\ \eta')$ is now possible using these amplitudes.  For
the amplitude solutions (A1, A2) preferred by fits to singly-Cabibbo-suppressed
decays, we find ${\cal B}(D_s \to \rho^+ \eta) = (5.6 \pm 1.2,~5.55 \pm 0.60)\%$,
to be compared with the experimental value of $(6.63 \pm 0.56)\%$, and ${\cal
B}(D_s \to \rho^+ \eta') = (2.9 \pm 0.3,~1.89 \pm 0.20)\%$, to be compared with
the experimental value of $(12.5 \pm 1.0)\%$.  The agreement between prediction
and experiment for ${\cal B}(D_s \to \rho^+ \eta)$ is good for the solutions
A1, A2, B1, and B2, while no solution gives agreement for ${\cal B}(D_s \to
\rho^+ \eta')$.  We await forthcoming CLEO data on this mode.

\begin{table}
\caption{Solution for annihilation amplitudes in Cabibbo-favored charmed meson
decays to $PV$ final states.
\label{tab:cfampsd}}
\begin{center}
\begin{tabular}{c c c c r c} \hline \hline
No.&   $PV$  & Magnitude &  Relative  & \multicolumn{2}{c}{Prediction}  \\
   &amplitude&($10^{-6}$)&   phase    & \multicolumn{2}{c}{($\%$)} \\ \hline
A1$^a$ &  $A_P$  &$1.36_{-1.04}^{+1.16}$&$\delta_{A_P} = (-151^{+83}_{-74})^\circ$
       &${\cal{B}}(D_s^+ \to \eta\, \rho^+) =$&$5.6 \pm 1.2$\\
       &  $A_V$  &$1.25_{-0.31}^{+0.34}$&$\delta_{A_V} = (-19^{+10}_{-9})^\circ$
       &${\cal{B}}(D_s^+ \to \eta'\,\rho^+)=$&$2.9 \pm 0.3$\\ \hline
A2$^b$ &  $A_P$  &$2.15_{-0.18}^{+0.22}$&$\delta_{A_P} = (-179^{+32}_{-9})^\circ$
       &${\cal{B}}(D_s^+ \to \eta\, \rho^+) =$&$5.55 \pm 0.60$\\
       &  $A_V$  &$1.23_{-0.19}^{+0.31}$&$\delta_{A_V} = (-19^{+34}_{-14})^\circ$
       &${\cal{B}}(D_s^+ \to \eta'\,\rho^+) =$&$1.89 \pm 0.20$\\ \hline
 A3    &  $A_P$  &$1.24_{-0.24}^{+0.34}$&$\delta_{A_P} = (-89^{+10}_{-14})^\circ$
       &${\cal{B}}(D_s^+ \to \eta\, \rho^+) =$&$4.20 \pm 0.81$\\
       &  $A_V$  &$0.96_{-0.22}^{+0.27}$&$\delta_{A_V} = ( 34^{+21}_{-14})^\circ$
       &${\cal{B}}(D_s^+ \to \eta'\,\rho^+) =$&$1.45 \pm 0.28$\\ \hline
 A4    &  $A_P$  &$4.27_{-0.21}^{+0.42}$&$\delta_{A_P} = (-109^{+14}_{-5})^\circ$
       &${\cal{B}}(D_s^+ \to \eta\, \rho^+) =$&$2.77 \pm 0.27$\\
       &  $A_V$  &$3.20_{-0.19}^{+0.23}$&$\delta_{A_V} = (+72^{+6}_{-4})^\circ$
       &${\cal{B}}(D_s^+ \to \eta'\,\rho^+) =$&$1.77 \pm 0.18$\\ \hline
 A5    &  $A_P$  &$2.88_{-0.24}^{+0.35}$&$\delta_{A_P} = (-123^{+6}_{-4})^\circ$
       &${\cal{B}}(D_s^+ \to \eta\, \rho^+) =$&$0.58 \pm 0.06$\\
       &  $A_V$  &$1.93_{-0.27}^{+1.21}$&$\delta_{A_V} = (69^{+15}_{-5})^\circ$
       &${\cal{B}}(D_s^+ \to \eta'\,\rho^+) =$&$0.70 \pm 0.07$\\ \hline
 A6    &  $A_P$  &$2.88_{-0.31}^{+0.22}$&$\delta_{A_P} = (+122^{+5}_{-6})^\circ$
       &${\cal{B}}(D_s^+ \to \eta\, \rho^+) =$&$1.61 \pm 0.17$\\
       &  $A_V$  &$2.85_{-0.26}^{+0.21}$&$\delta_{A_V} = (-36 \pm 7)^\circ$
       &${\cal{B}}(D_s^+ \to \eta'\,\rho^+) =$&$0.43 \pm 0.04$\\ \hline\hline
 B1    &  $A_P$  &$1.57_{-0.32}^{+0.82}$&$\delta_{A_P} = (+121^{+19}_{-9})^\circ$
       &${\cal{B}}(D_s^+ \to \eta\, \rho^+) =$&$7.08 \pm 1.03$\\
       &  $A_V$  &$1.74_{-0.28}^{+0.44}$&$\delta_{A_V} =(-96^{+7}_{-6})^\circ$
       &${\cal{B}}(D_s^+ \to \eta'\,\rho^+) =$&$2.53 \pm 0.37$\\ \hline
 B2    &  $A_P$  &$1.35_{-0.27}^{+0.51}$&$\delta_{A_P} = (- 74^{+12}_{-9})^\circ$
       &${\cal{B}}(D_s^+ \to \eta\, \rho^+) =$&$5.38_{-2.11}^{+2.03}$\\
       &  $A_V$  &$1.52_{-0.21}^{+0.70}$&$\delta_{A_V} = (+150^{+44}_{-10})^\circ$
       &${\cal{B}}(D_s^+ \to \eta'\,\rho^+) =$&$1.86_{-0.73}^{+0.70}$\\ \hline
 B3    &  $A_P$  &$3.85_{-0.24}^{+0.39}$&$\delta_{A_P} = (+111^{+14}_{-5})^\circ$
       &${\cal{B}}(D_s^+ \to \eta\, \rho^+) =$&$2.42 \pm 0.16$\\
       &  $A_V$  &$2.78_{-0.22}^{+0.37}$&$\delta_{A_V} = (-68^{+17}_{-7})^\circ$
       &${\cal{B}}(D_s^+ \to \eta'\,\rho^+) =$&$1.01 \pm 0.07$\\ \hline
 B4    &  $A_P$  &$1.74_{-0.23}^{+0.34}$&$\delta_{A_P} = (+ 77^{+41}_{-10})^\circ$
       &${\cal{B}}(D_s^+ \to \eta\, \rho^+) =$&$3.04 \pm 0.70$\\
       &  $A_V$  &$1.16_{-0.23}^{+0.27}$&$\delta_{A_V} = (-140 \pm 12)^\circ$
       &${\cal{B}}(D_s^+ \to \eta'\,\rho^+) =$&$1.18 \pm 0.27$\\ \hline
 B5    &  $A_P$  &$4.12_{-0.31}^{+0.24}$&$\delta_{A_P} = (+111^{+6}_{- 9})^\circ$
       &${\cal{B}}(D_s^+ \to \eta\, \rho^+) =$&$1.30 \pm 0.10$\\
       &  $A_V$  &$3.22_{-0.38}^{+0.29}$&$\delta_{A_V} = (-60^{+8}_{-11})^\circ$
       &${\cal{B}}(D_s^+ \to \eta'\,\rho^+) =$&$0.571_{-0.044}^{+0.045}$\\ \hline
 B6    &  $A_P$  &$0.67_{-0.29}^{+0.26}$&$\delta_{A_P} = (+45^{+22}_{-25})^\circ$
       &${\cal{B}}(D_s^+ \to \eta\, \rho^+) =$&$4.80 \pm 2.54$\\
       &  $A_V$  &$1.28_{-0.20}^{+0.23}$&$\delta_{A_V} = (+168^{+11}_{-15})^\circ$
       &${\cal{B}}(D_s^+ \to \eta'\,\rho^+)=$&$3.42 \pm 1.81$\\ \hline\hline
\end{tabular}
\end{center}
\leftline{$^a$Preferred solution based on fit to singly-Cabibbo-suppressed
decays.}
\leftline{$^b$Alternative solution giving acceptable fit to
singly-Cabibbo-suppressed decays.}

\end{table}

\section{SINGLY-CABIBBO-SUPPRESSED DECAYS}

The topological amplitude decomposition of singly-Cabibbo-suppressed decays of
$D^0 \to PV$ is listed in Table \ref{tab:SCSPVa} along with the measured
branching ratios and amplitudes for the decays.  Unlike the $D \to PP$ case
\cite{Bhattacharya:2008ss}, here we have neglected the Okubo-Zweig-Iizuka
(OZI) suppressed disconnected diagrams that form the Singlet-Exchange $(SE\,')$
and Singlet-Annihilation $(SA\,')$ amplitudes.

\begin{table}
\caption{Branching ratios and invariant amplitudes for singly-Cabibbo-suppressed
decays of charmed mesons to one pseudoscalar and one vector meson.
\label{tab:SCSPVa}}
\begin{center}
\begin{tabular}{c l c c c c}
\hline \hline
Meson & Decay & Representation
     & ${\cal B}$ \cite{Amsler:2008} & $p^*$ & $|{\cal A}|$ \\
 & mode & & ($\%$) & (MeV) & $(10^{-6})$ \\ \hline \hline
$D^0$ & $\pi^+\, \rho^-$& $-(T_V\,' + E_P\,')$ & 0.497$\pm$0.023 & 763.8
 & 1.25$\pm$0.03 \\
 & $\pi^-\, \rho^+$ & $-(T_P\,' + E_V\,')$ & 0.980$\pm$0.040 & 763.8
 & 1.76$\pm$0.04 \\
 & $\pi^0\, \rho^0$ & $\frac{1}{2}(E_P\,' + E_V\,' - C_P\,' - C_V\,')$
 & 0.373$\pm$0.022 & 764.2 & 1.08$\pm$0.03 \\
 & $K^+\, K^{*-}$ & $T_V\,' + E_P\,'$ & 0.153$\pm$0.015 & 609.8
 & 0.97$\pm$0.05 \\
 & $K^-\, K^{*+}$ & $T_P\,' + E_V\,'$ & 0.441$\pm$0.021 & 609.8
 & 1.65$\pm$0.04 \\
 & $K^0\, \ol{K}^{*0}$ & $E_V\,' - E_P\,'$ & $< 0.18$ & 605.3 & \\
 & $\ol{K}^{0}\, K^{*0}$ & $E_P\,' - E_V\,'$ & $< 0.09$ & 605.3 & \\
 & $\pi^0\, \phi$ & $\frac{1}{\sqrt{2}} C_P\,'$ & 0.124$\pm$0.012& 644.7 & 0.81$\pm$0.04\\
 & $\pi^0\, \omega$ & $\frac{1}{2}(E_P\,' + E_V\,' - C_P\,' + C_V\,')$
 & & 761.2 & \\
 & $\eta\, \rho^0$ & $\frac{1}{\sqrt{6}}(2 C_V\,' - C_P\,' - E_P\,' - E_V\,')$
 & & 652.0 & \\
 & $\eta\, \omega$ & $- \frac{1}{\sqrt{6}}(2 C_V\,' + C_P\,' + E_P\,' +E_V\,')$
 & & 488.8 & \\
 & $\eta\, \phi$ & $\frac{1}{\sqrt{3}}(C_P\,' - E_P\,' - E_V\,')$ & & 648.1
 & \\
 & $\eta\,' \rho^0$ & $\frac{1}{2 \sqrt{3}}(E_P\,' + E_V\,' + C_P\,' +C_V\,')$
 & & 342.5 & \\
 & $\eta\,' \omega$ & $\frac{1}{2 \sqrt{3}}(E_P\,' + E_V\,' + C_P\,' - C_V\,')$
 & & 333.5 & \\ \hline
$D^+$  &$\rho^0\, \pi^+$   &$\frac{1}{\s}(A_P\,'-A_V\,'-C_P\,'-T_V\,')$   &0.082$\pm$0.015&$767$&0.32$\pm$0.03\\
       &$\omega\, \pi^+$   &$-\frac{1}{\s}(A_P\,'+A_V\,'+C_P\,'+T_V\,')$  &$<0.034$&$764$       &\\
       &$\phi\, \pi^+$     &$C_P\,'$                                      &0.620$\pm$0.070&$647$&1.13$\pm$0.06\\
       &$\ol{K}^{*0}\, K^+$&$(T_V\,'-A_V\,')$                             &0.435$\pm$0.048&$611$&1.03$\pm$0.06\\
       &$\pi^0\, \rho^+$   &$\frac{1}{\s}(A_V\,'-A_P\,'-C_V\,'-T_P\,')$   &&$767$               &\\
       &$\eta\, \rho^+$    &$\frac{1}{\sx}(A_V\,'+A_P\,'+2C_V\,'+T_P\,')$ &$<0.7$&$656$         &\\
       &$\eta\,' \rho^+$   &$\frac{1}{\sx}(C_V\,'-A_V\,'-A_P\,'-T_P\,')$  &$<0.5$&$349$         &\\
       &$\ol{K}^0\, K^{*+}$&$(T_P\,'-A_P\,')$                             &3.18$\pm$1.38&$612$  &2.78$\pm$0.60\\ \hline
$D_s^+$&$\pi^+\, K^{*0}$   &$(A_V\,'-T_V\,')$                             &0.225$\pm$0.039&$773$&0.79$\pm$0.07\\
       &$\pi^0\, K^{*+}$   &$-\frac{1}{\s}(C_V\,'+A_V\,')$                &&$775$&\\
       &$\eta\,  K^{*+}$   &$\frac{1}{\st}(T_P\,'+2C_V\,'+A_P\,'-A_V\,')$ &&$661$&\\
       &$\eta\,' K^{*+}$   &$\frac{1}{\sx}(2T_P\,'+C_V\,'+2A_P\,'+A_V\,')$&&$337$&\\
       &$K^0\,   \rho^+$   &$(A_P\,'-T_P\,')$                             &&$743$&\\
       &$K^+\, \rho^0$     &$-\frac{1}{\s}(C_P\,'+A_P\,')$                &0.27$\pm$0.05&$745$&0.92$\pm$0.09\\
       &$K^+\, \omega$     &$-\frac{1}{\s}(C_P\,'-A_P\,')$                &&$741$&\\
       &$K^+\, \phi$       &$T_V\,'+C_P\,'+A_V\,'$                        &$<0.057$&$607$&\\ \hline \hline

\end{tabular}
\end{center}
\end{table}

\begin{table}
\caption{Global $\chi^2$ values for fits to singly-Cabibbo-suppressed $D \to
PV$ decays. Also included are the process that contribute the most to a high
$\chi^2$ value.}
\label{tab:scschisq}
\begin{center}
\begin{tabular}{l c l c c c} \hline \hline
   No.  & Global &  \multicolumn{4}{c}{Worst Processes (High $\Delta\chi^2$ value)}\\
        &$\chi^2$&      Decay Channel       &${\cal{B}}_{th}(\%)$&${\cal{B}}_{expt}(\%)$&$\Delta\chi^2$\\\hline
 A1$^a$ &   61.8 &$D^+ \to \ol{K}^{*0}\,K^+$&   $0.17 \pm 0.04$  &   $0.435 \pm 0.048$  &  16.1 \\
        &        &$D^+ \to \omega\,\pi^+$   &   $0.16 \pm 0.04$  &        $<0.034$      &  11.6 \\\hline
 A2$^b$ &   65.9 &$D^+ \to \ol{K}^{*0}\,K^+$&   $0.17 \pm 0.03$  &   $0.435 \pm 0.048$  &  21.4 \\
        &        &$D^0 \to \rho^0\,\pi^0$   &   $0.27 \pm 0.02$  &   $0.373\pm
0.022$  &  10.1 \\\hline
 A3     &  341.4 &$D^0 \to \rho^0\,\pi^0$   &$(4.3\pm3.1)\times10^{-3}$& $0.373\pm0.022$& 275.2 \\
        &        &$D^+ \to \rho^0\,\pi^+$   &$(1.5\pm4.0)\times10^{-3}$& $0.082\pm0.015$&  25.1 \\\hline
 A4     &  167.1 &$D^0 \to \rho^0\,\pi^0$   &   $0.12 \pm 0.01$  &   $0.373 \pm 0.022$  &  95.4 \\
        &        &$D^+ \to \rho^0\,\pi^+$   &   $0.73 \pm 0.12$  &   $0.082 \pm 0.015$  &  31.4 \\\hline
 A5     &  324.1 &$D^0 \to \rho^0\,\pi^0$   &$(6.1\pm3.1)\times10^{-3}$& $0.373\pm0.022$& 272.6 \\
        &        &$D^+ \to K^{*0}\,\ol{K}^0$&   $0.19 \pm 0.02$  &   $<0.09$            &  11.9 \\\hline
 A6     &  149.8 &$D^+ \to \rho^0\,\pi^+$   &   $0.91 \pm 0.09$  &   $0.082 \pm 0.015$  &  51.1 \\
        &        &$D^+ \to \ol{K}^{*0}\,K^+$&   $0.12 \pm 0.03$  &   $0.435 \pm 0.048$  &  32.1 \\\hline\hline
 B1     &  244.0 &$D^0 \to \rho^0\,\pi^0$   &   $0.12 \pm 0.01$  &   $0.373 \pm 0.022$  &  95.3 \\
        &        &$D^0 \to \phi\,\pi^0$     &   $0.042\pm0.003$  &   $0.124 \pm 0.012$  &  45.3 \\\hline
 B2     &  155.7 &$D^0 \to \phi\,\pi^0$     &   $0.042\pm0.003$  &   $0.124 \pm 0.012$  &  45.3 \\
        &        &$D^+ \to \phi\,\pi^+$     &   $0.21 \pm 0.01$  &   $0.62  \pm 0.07$   &  32.9 \\\hline
 B3     &  165.7 &$D^0 \to \phi\,\pi^0$     &   $0.042\pm0.003$  &   $0.124 \pm 0.012$  &  45.3 \\
        &        &$D^+ \to \phi\,\pi^+$     &   $0.21 \pm 0.01$  &   $0.62  \pm 0.07$   &  32.9\\\hline
 B4     &  151.7 &$D^0 \to \phi\,\pi^0$     &   $0.042\pm0.002$  &   $0.124 \pm 0.012$  &  45.3 \\
        &        &$D^+ \to \rho^0\,\pi^+$   &   $1.44 \pm 0.23$  &   $0.082 \pm 0.015$  &  34.4 \\\hline
 B5     &  518.8 &$D^0 \to \rho^0\,\pi^0$   &$(5.4\pm2.8)\times10^{-3}$& $0.373\pm0.022$& 274.8 \\
        &        &$D^+ \to \rho^0\,\pi^+$   &   $1.71 \pm 0.21$  &   $0.082 \pm 0.015$  &  59.3 \\\hline
 B6     &  401.3 &$D^0 \to \rho^0\,\pi^0$   &   $0.015\pm0.006$  &   $0.373 \pm 0.022$  & 245.9 \\
        &        &$D^0 \to \phi\,\pi^0$     &   $0.042\pm0.003$  &   $0.124 \pm 0.012$  &  45.3 \\\hline\hline
\end{tabular}
\end{center}
\leftline{$^a$Preferred solution.  $^b$Alternative solution.}

\end{table}

We now make use of the amplitudes determined in Section III to predict
the singly-Cabibbo-suppressed decay amplitudes.  Here we assume the simple
hierarchy of amplitudes explained in Section II. Based on the available
data  we calculated the global $\chi^2$ of singly-Cabibbo-suppressed
$D \to PV$ decays for solutions A1--A6 and B1--B6. Solutions A1 and A2
have the two lowest values of $\chi^2$ and hence were chosen as the
preferred and alternative solutions. Table \ref{tab:scschisq} summarizes
the global $\chi^2$ values for each of the twelve solutions. It also
includes, for each solution, two processes that contribute the most
towards a high value of $\chi^2$.

One notes in Table \ref{tab:scschisq} that the main processes contributing
to high global $\chi^2$ for all solutions are $D^0 \to \phi \pi^0$
and $D^0 \to \rho^0 \pi^0$. The solutions B1-6, which correspond to
$|C_P| < |T_V|$, yield high $\chi^2$ for the process $D^0 \to \phi \pi^0$.
The amplitude of this process depends only on $C_P\,'$. This shows that
$|C_P| < |T_V|$ is not favored by the process $D^0 \to \phi \pi^0$. The
processes $D^0 \to \rho^0 \pi^0$ and $D^+ \to \rho^0 \pi^+$ contribute to high
$\chi^2$ for the solutions A3-6.

The predicted and experimental $D^0$ branching ratios are in qualitative
agreement but with some notable exceptions.  The predictions for $D^0 \to \pi
\rho$ fall slightly short of experiment for all charge states, most prominently
for $\pi^0 \rho^0$.  Recall that the predicted branching ratio for $D^0 \to
\pi^+ \pi^-$ lies significantly {\it above} the experimental value
\cite{Bhattacharya:2008ss}.  The predictions for $D^0 \to K^+ K^{*-}$ and
$D^0 \to K^- K^{*+}$ are not badly obeyed, while those for $D^0 \to K^0
\overline{K}^{*0}$ and $D^0 \to \overline{K}^0 K^0$ are far below the
current experimental upper limits.  The predicted branching ratio for
$D^0 \to \pi^0 \phi$ is approximately same as the observed value. The value of
$\chi^2$ for solutions \# A1 and A2 are respectively 61.8 and 65.9 (Table
\ref{tab:scschisq}), where we have used the 18 data points for which the
branching ratios are available.

In Table \ref{tab:SCSPVb} we present our predictions for branching ratios of
singly-Cabibbo-suppressed $D \to PV$ modes corresponding to the two solutions
A1 and A2 having the lowest value of global $\chi^2$ for these modes.  There is
little one can do to distinguish between them given the available data on
branching ratios.  Both solutions yield fairly similar central values for most
of the singly-Cabibbo-suppressed $D \to PV$ modes.  A slight
distinction may be made in a few cases.  For example, the predicted central
values of $\b(D^0 \to (K^0\,\ol{K}^{*0}, \ol{K}^0\,K^{*0}))$ are larger for
solution A1 than for A2, though differing only by $1.5 \sigma$.  Another
example is the process $D^0 \to \pi^0 \omega$, for which the central value
of the branching ratio in solution A2 is nearly three times its value in A1.
Still another example is the process $D^+ \to \eta' \rho^+$, for which the
predicted (very small) branching ratio in A1 is twice its value in A2.
Measurements of the branching ratios for both Cabibbo-favored and
singly-Cabibbo-suppressed decays with higher precision will be necessary in
order to distinguish between the two solutions.

\begin{table}
\caption{Comparison between predicted amplitudes based on Cabibbo-favored decays
and the experimental values for singly-Cabibbo-suppressed decays of $D^0$ to a
pseudoscalar and a vector meson.  Predictions are listed for preferred (A1)
and alternative (A2) solutions.
\label{tab:SCSPVb}}
\begin{center}
\begin{tabular}{c c c c c}
\hline \hline
       & PV Decay           & Experimental     & \multicolumn{2}{c}{Predicted ${\cal{B}}~(\%)$}  \\
Meson  & Mode               & ${\cal{B}}~(\%)$ &  Solution A1    & Solution A2\\ \hline \hline
$D^0$  & $\pi^+\, \rho^-$   & $0.497\pm0.023$  & $0.39\pm0.03$     & $0.39\pm0.03$     \\
       & $\pi^-\, \rho^+$   & $0.980\pm0.040$  & $0.84\pm0.06$     & $0.84\pm0.06$     \\
       & $\pi^0\, \rho^0$   & $0.373\pm0.022$  & $0.29\pm0.02$     & $0.27\pm0.02$     \\
       & $K^+\, K^{*-}$     & $0.153\pm0.015$  & $0.20\pm0.01$     & $0.20\pm0.01$     \\
       & $K^-\, K^{*+}$     & $0.441\pm0.021$  & $0.43\pm0.03$     & $0.43\pm0.03$     \\
       & $K^0\, \ol{K}^{*0}$& $< 0.18$         & $0.0080\pm0.0036$ & $0.0020\pm0.0016$ \\
       & $\ol{K}^0\, K^{*0}$& $< 0.09$         & $0.0080\pm0.0036$ & $0.0020\pm0.0016$ \\
       & $\pi^0\, \phi$     & $0.124\pm0.012$  & $0.122\pm0.007$   & $0.122\pm0.007$   \\
       & $\pi^0\, \omega$   &                  & $0.043\pm0.008$   & $0.119\pm0.012$   \\
       & $\eta\, \rho^0$    &                  & $0.106\pm0.013$   & $0.095\pm0.010$   \\
       & $\eta\, \omega$    &                  & $0.140\pm0.009$   & $0.127\pm0.009$   \\
       & $\eta\, \phi$      &                  & $0.093\pm0.009$   & $0.14\pm0.01$     \\
       & $\eta\,' \rho^0$   &                  & $0.0154\pm0.0009$ & $0.0158\pm0.0009$ \\
       & $\eta\,' \omega$   &                  & $0.0066\pm0.0005$ & $0.0077\pm0.0005$ \\ \hline
$D^+$  &$\rho^0\, \pi^+$    & $0.082\pm0.015$  & $0.097\pm0.048$   & $0.23\pm0.12$     \\
       &$\omega\, \pi^+$    & $<0.034$         & $0.15\pm0.04$     & $0.14\pm0.12$     \\
       &$\phi\, \pi^+$      & $0.620\pm0.070$  & $0.62\pm0.04$     & $0.62\pm0.04$     \\
       &$\ol{K}^{*0}\, K^+$ & $0.435\pm0.048$  & $0.17\pm0.04$     & $0.17\pm0.03$     \\
       &$\pi^0\, \rho^+$    &                  & $0.062\pm0.047$   & $0.012\pm0.015$   \\
       &$\eta\, \rho^+$     &     $<0.7$       & $0.0017\pm0.0040$ & $0.0057\pm0.013$  \\
       &$\eta\,' \rho^+$    & $<0.5$           & $0.083\pm0.010$   & $0.044\pm0.005$   \\
       &$\ol{K}^0\, K^{*+}$ & $3.18\pm1.38$    & $1.66\pm0.20$     & $1.66\pm0.12$     \\ \hline
$D_s^+$&$\pi^+\, K^{*0}$    & $0.225\pm0.039$  & $0.15\pm0.04$     & $0.15\pm0.03$     \\
       &$\pi^0\, K^{*+}$    &                  & $0.049\pm0.012$   & $0.020\pm0.008$   \\
       &$\eta\,  K^{*+}$    &                  & $0.014\pm0.011$   & $0.012\pm0.008$   \\
       &$\eta\,' K^{*+}$    &                  & $0.029\pm0.006$   & $0.015\pm0.003$   \\
       &$K^0\,   \rho^+$    &                  & $1.29\pm0.15$     & $1.29\pm0.09$     \\
       &$K^+\, \rho^0$      & $0.27\pm0.05$    & $0.33\pm0.05$     & $0.42\pm0.05$     \\
       &$K^+\, \omega$      &                  & $0.108\pm0.029$   & $0.072\pm0.033$   \\
       &$K^+\, \phi$        & $<0.057$         & $0.038\pm0.009$   & $0.037\pm0.028$   \\ \hline \hline
\end{tabular}
\end{center}
\end{table}

\section{DOUBLY-CABIBBO-SUPPRESSED DECAYS}

We now characterize the doubly-Cabibbo-suppressed or wrong-sign (WS) decays of
$D \to PV$.  A detailed list of possible decays and the corresponding
topological amplitude decompositions are given in Table \ref{tab:WSPV}. We
used the Cabibbo-favored amplitudes calculated in section III to
predict the WS amplitudes, using the simple hierarchy of amplitudes as
explained in Section II.  The predicted amplitudes have been included in Table
\ref{tab:WSPV} for the preferred (A1) and alternative (A2) solutions.

The experimental values for the following decays are available in the
literature \cite{Amsler:2008}:
\bea
{\cal{B}}(D^0 \to K^{*+}\, \pi^-) &=& (3.0^{+3.9}_{-1.2}) \times 10^{-4} \\
{\cal{B}}(D^+ \to K^{*0}\, \pi^+) &=& (4.35 \pm 0.9) \times 10^{-4}
\eea
The predicted values for these branching ratios (Table \ref{tab:WSPV})
are in satisfactory agreement with the experimental values quoted above.  An
interesting point to note is that both solutions A1 and A2 give the same
predicted central values for these branching ratios, but A2 has a larger error
bar on both of them.  Several other branching ratios in Table \ref{tab:WSPV}
predicted to exceed $10^{-4}$ may help to distinguish between solutions A1
and A2.  These include $\b(D^0 \to K^{*0} \pi^0)$, $\b(D^+ \to K^{*+} \pi^0)$,
and $\b(D^+ \to K^+ \rho^0)$.  Reduction in errors on predictions will be
needed in order that these distinctions exceed $2$--$2.5 \sigma$.  Some of the
doubly-Cabibbo-suppressed decays in Table \ref{tab:WSPV} may be observable
in Dalitz plots of $D$ decays to three pseudoscalars through interference with
Cabibbo-favored $PV$ decays.  For example, $D^0 \to K_S \pi^+ \pi^-$ might be
able to provide new information about the decay process $D^0 \to K^{*+} \pi^-$,
while $D^+ \to K_S \pi^+ \pi^0$ could provide information about $D^+ \to K^{*+}
\pi^0$.

\begin{table}
\caption{Branching ratios and invariant amplitudes for doubly-Cabibbo-%
suppressed decays of charmed mesons to one pseudoscalar and one vector meson.
Predictions are shown for favored (A1) and alternative (A2) solutions.
\label{tab:WSPV}}
\begin{center}
\begin{tabular} {c l c c c c}
\hline \hline
Meson  & Decay & Representation &$p^*$& \multicolumn{2}{c}{Predicted ${\cal{B}} (10^{-4})$}  \\
       &  mode  &           &(MeV)& Solution A1 & Solution A2 \\ \hline \hline
$D^0$  &$K^{*+}\, \pi^-$  &$T_P\, '' + E_V\,''$                 &$711$&$3.63\pm0.26$    &$3.63\pm0.27$     \\
       &$K^{*0}\, \pi^0$  &$(C_P\,'' - E_V\,'')/\s$             &$709$&$0.55\pm0.06$    &$0.80\pm0.08$     \\
       &$K^{*0}\, \eta$   &$(C_P\,'' - E_P\,'' + E_V\,'')/\st$  &$580$&$0.38\pm0.04$    &$0.37\pm0.04$     \\
       &$K^{*0}\, \eta\,'$&$-(C_P\,'' + 2E_P\,'' + E_V\,'')/\sx$&$102$&$0.0046\pm0.0004$&$0.0052\pm0.0004$ \\
       &$K^+\, \rho^{-}$  &$T_V\,'' + E_P\,''$                  &$675$&$1.46\pm0.10$    &$1.46\pm0.10$     \\
       &$K^0\, \rho^{0}$  &$(C_V\,'' - E_P\,'')/\s$             &$674$&$0.70\pm0.07$    &$0.39\pm0.05$     \\
       &$K^0\, \omega$    &$-(C_V\,'' + E_P\,'')/\s$            &$670$&$0.58\pm0.06$    &$0.52\pm0.06$     \\
       &$K^0\, \phi$      &$- E_V\,''$                          &$521$&$0.16\pm0.03$    &$0.33\pm0.03$     \\ \hline
$D^+$  &$K^{*0}\, \pi^+$  &$C_P\,'' + A_V\,''$                  &$712$&$2.94\pm0.52$    &$2.94\pm0.65$     \\
       &$K^{*+}\, \pi^0$  &$(T_P\,'' - A_V\,'')/\s$             &$714$&$3.74\pm0.49$    &$2.71\pm0.30$     \\
       &$K^{*+}\, \eta$   &$-(T_P\,'' - A_P\,'' + A_V\,'')/\st$ &$586$&$3.37\pm0.43$    &$3.37\pm0.25$     \\
       &$K^{*+}\, \eta\,'$&$(T_P\,'' + 2A_P\,'' + A_V\,'')/\sx$ &$137$&$0.0095\pm0.0029$&$0.0026\pm0.0010$ \\
       &$K^0\, \rho^+$    &$C_V\,'' + A_P\,''$                  &$677$&$3.43\pm0.75$    &$3.43\pm0.47$     \\
       &$K^+\, \rho^0$    &$(T_V\,'' - A_P\,'')/\s$             &$679$&$2.17\pm0.40$    &$3.01\pm0.24$     \\
       &$K^+\, \omega$    &$(T_V\,'' + A_P\,'')/\s$             &$675$&$0.64\pm0.21$    &$0.26\pm0.07$     \\
       &$K^+\, \phi$      &$A_V\,''$                            &$527$&$0.12\pm0.06$    &$0.12\pm0.06$     \\ \hline
$D_s^+$&$K^{*0}\, K^+$    &$T_V\,'' + C_P\,''$                  &$682$&$0.20\pm0.02$    &$0.20\pm0.02$     \\
       &$K^{*+}\, K^0$    &$T_P\,'' + C_V\,''$                  &$683$&$1.18\pm0.16$    &$1.18\pm0.18$     \\
\hline \hline
\end{tabular}
\end{center}
\end{table}





\section{FACTORIZATION COMPARISONS}

In the current section we compare our results for the amplitudes of $T_P$
and $T_V$ with the values extracted from explicit evaluation of the tree
diagram assuming factorization \cite{Bjorken:1990}. In order to calculate
$T_P$ we use the decay $D^0 \to K^-\, \rho^+$. In this scenario the
spectator $\ol{u}$ quark goes from $D^0$ to the pseudoscalar $K^-$ and so
we use the standard form of the $(D \to P)$ current \cite{Wisgur:1990}:
\beq
H_\mu = f_+(q^2) (p_D + p_K)_\mu - f_-(q^2) (p_D - p_K)_\mu
\eeq
where $f_+$ and $f_-$ are the relevant form factors. The current we use for
the vector meson is \cite{Rosner:1990}:
\beq
\rho^{\mu} = \epsilon^{\mu} m_{\rho} f_{\rho}
\eeq
where $\epsilon^{\mu}$ represents the polarization of the vector meson,
$m_{\rho}$ is its mass and $f_{\rho}$ is the associated decay constant. The
invariant amplitude and the decay rate for the process $D^0 \to K^-\, \rho^+$
via the tree diagram may then be written as
\bea
{\cal{M}}(D^0 \to K^-\,\rho^+)_{T_P}&=&-{\it{i}}\,\frac{G_F}{\sqrt{2}}\,V_{cs}
\,V_{ud}^{*}H_{\mu}\,\rho^{\mu}\\
\Gamma(D^0 \to K^-\,\rho^+)_{T_P}&=&\frac{p^*}{8 \pi M_{D^0}^2}\sum_{\epsilon_
{\mu}q^{\mu}=0}|{\cal{M}}(D^0 \to K^-\, \rho^+)_{T_P}|^2
\eea
After summing over the $\rho$ polarization and taking the modulus squared of
the invariant amplitude one obtains the final form for $|T_P|$:
\bea
|T_P|&=&\frac{G_F}{\sqrt{2}} \frac{|V_{ud}||V_{cs}||f_{+}(m_{\rho}^2)|f_{\rho}}{p^*}\\
&&\times\sqrt{(m_D^2 - m_K^2)^2 - m_{\rho}^2 (m_D^2 + m_K^2 + 2 m_D \sqrt{m_K^2 + p^{*2}})}\\
&=& (5.45 \pm 0.07) \times 10^{-6}
\eea
which is to be compared with the values quoted in Table \ref{tab:cfampsc},
and favors solution A2 over A1.

In obtaining the result stated above we used $|f_+(m_\rho^2)||V_{cs}| = 0.869
\pm 0.009$ \cite{Shipsey:2008}. The particle masses and the quantity $|V_{ud}|$
were taken from \cite{Amsler:2008}. $p^*$ is as quoted in Table \ref{tab:CFPV}.
We calculated the value of $f_\rho$ using the following formula:
\bea
f_\rho&=&f_\pi {\left[\frac{{\cal{B}}(\tau^- \to \nu_{\tau}\,\rho^-)}{{\cal{B}}(\tau^- \to \nu_{\tau}\,\pi^-)}\right]}^{\frac{1}{2}}
\frac{m_{\tau}^2 - m_{\pi}^2}{m_{\tau}^2 - m_{\rho}^2}
\frac{m_{\tau}}{\sqrt{m_{\tau}^2 + 2 m_{\rho}^2}}\\
&=&(209 \pm 1.6){\rm~MeV}
\eea
where once again the particle masses and branching fractions were taken from
\cite{Amsler:2008}.

A similar approach may be taken in order to evaluate $|T_V|$ by looking at the
decay $D^0 \to K^{*-}\, \pi^+$ via the tree diagram. In this case the spectator
$\ol{u}$ quark goes from $D^0$ to the vector meson $K^{*-}$, so we use the
standard forms of the $(D \to V)$ vector and axial-vector currents
\cite{Wisgur:1990} and the pion current \cite{Rosner:1990}:
\bea
V_{\mu}&=&\it{i}g\epsilon_{\mu\rho\tau\sigma}\epsilon^{*\rho}(p_D + p_{K^*})^{\sigma}(p_D - p_{K^*})^{\tau}\\
A_{\mu}&=&f\epsilon^*_{\mu} + a_+(\epsilon^*{\cdot}p_D)(p_D + p_{K^*})_{\mu} + a_-(\epsilon^*{\cdot}p_D)(p_D - p_{K^*})_{\mu}\\
\pi^{\mu}&=&\it{i}f_{\pi}q^{\mu}
\eea
We obtain for the amplitude $|T_V|$ the following expression:
\beq
|T_V|=\frac{G_F}{\sqrt{2}}|V_{cs}||V_{ud}|f_{\pi}\frac{m_D}{m_{K^*}}
|f + a_+ (m_D^2 - m_{K^*}^2) + a_- m_{\pi}^2|
\eeq
In principle this can be used to calculate $T_V$ once the form factors are
given.  However, we may adopt a simplification using a result from
Ref.\ \cite{Rosner:1999}, based on the earlier discussion in Ref.\
\cite{Rosner:1990}.  In the heavy-quark limit one expects $\Gamma(D \to
\bar K^* \pi^+)_T = \Gamma(D \to\bar K \pi^+)_T$ and hence
\beq
p^*_{K \pi} |T|^2_{K \pi} = (p^*_{K^* \pi})^3 |T|^2_{K^* \pi}~,
\eeq
where $T_{K^* \pi} = T_V$.  In Ref.\ \cite{Bhattacharya:2008ss} we found in a
fit to $D \to PP$ amplitudes that $|T|_{K \pi} = (2.78 \pm 0.13) \times
10^{-6}$ GeV.  With $p^*_{K \pi} = 0.861$ GeV and $p^*_{K^* \pi} = 0.711$ GeV we
then obtain the result
\beq
T_V = (4.3 \pm 0.2) \times 10^{-6}~,
\eeq
in reasonable agreement with the value of $(3.95 \pm 0.07) \times 10^{-6}$
quoted in Table \ref{tab:cfampsa}, especially considering the uncertainties
associated with QCD corrections and with the use of the heavy-quark limit
for the final strange quark.

\section{CONCLUSIONS}

We have used the flavor topology description to study the validity of flavor
SU(3) for describing $D \to PV$ decays, to obtain relative phases and
magnitudes of various contributing amplitudes, and to predict rates for
as-yet-unseen singly- and doubly-Cabibbo-suppressed decays.  We assumed
flavor $SU(3)$ to be an exact symmetry for the tree level diagrams.  We found
that singly-Cabibbo-suppressed decays favor a ratio of color-suppressed to tree
amplitudes $|C_P/T_V| > 1$, where the subscript denotes the meson ($P$ or $V$)
containing the spectator quark. The present data for the Cabibbo-favored decays
are compatible with twelve distinct sets of solutions for the amplitudes $T_P$,
$C_P$, $E_P$, $A_P$, $T_V$, $C_V$, $E_V$, and $A_V$ (up to discrete
ambiguities).  However, on the basis of experimental branching ratios for
singly-Cabibbo-suppressed decays we were able to choose two sets of solutions
giving substantially lower values for $\chi^2$ than the other ten.

Our predictions of the branching ratios for singly-Cabibbo-suppressed decays
deviate from the available experimental data in several cases, such as those in
the first four lines of Table VI. This shows that flavor $SU(3)$ is not an
exact symmetry. However flavor $SU(3)$ breaking, though present, is no worse
in $D \to PV$ decays than in the $D \to PP$ decays discussed in Ref.\
\cite{Bhattacharya:2008ss}.

Our prediction for the $D^+_s \to \eta\,'\rho^+$ branching ratio is much lower
than the available experimental value.  Either there are additional
contributions to $\eta'$ production which we have neglected, or the
experimental situation needs to be re-evaluated.

Our analysis of the singly-Cabibbo suppressed decays shows that processes such
as $D^0 \to \pi^0 \omega$ can be used to distinguish between the two most
likely amplitude solutions.  The mean values predicted for the branching ratios
of these processes differ by nearly a factor of three in the two solutions, but
experimental data are not yet available to resolve this problem.

The branching ratios predicted for doubly-Cabibbo-suppressed decays are close
to the experimental values in the two cases for which data are available.  A
precise measurement of a few of the other branching ratios may help select one
of the two most-favored amplitude solutions.

Finally, factorization computations of the tree amplitudes agree with results
obtained in direct analyses.  However, a more precise calculation of the
amplitudes using the factorization assumption could be done if data on the
relevant form factors were available.

\bigskip
\section*{ACKNOWLEDGMENTS}
This work was supported in part by the United States Department
of Energy through Grant No.\ DE FG02 90ER40560.

\end{document}